\documentclass[prl,twocolumn,showpacs]{revtex4}
\usepackage{tipa}
\usepackage{amssymb}
\usepackage{txfonts}
\usepackage{hyperref}
\usepackage{graphicx}
\usepackage{epsfig}
\begin{document}
\title{A position dependent atom-atom entanglement in real-time Cavity QED system}
\author{Yan-Qing Guo, Hai-Jing Cao, He-Shan Song\footnote{To whom the
correspondence should be addressed: hssong@dlut.edu.cn}}
\affiliation{Department of Physics, Dalian University of
Technology, Dalian, Liaoning 116023, P.R.China}

\pacs{03.67.Mn, 42.50.Pq}

\begin{abstract}
We study a special two-atom entanglement case in assumed Cavity
QED experiment in which only one atom effectively exchanges a
single photon with a cavity mode. We compute diatom entanglement
under position-dependent atomic resonant dipole-dipole interaction
(RDDI) for large interatomic separation limit. We show that the
RDDI, even which is much smaller than the maximal atomic Rabi
frequency, can induce distinct diatom entanglement. The peak
entanglement (PE) reaches a maximum when RDDI strength can compare
with the Rabi frequency of an atom.
\end{abstract}
\maketitle
\section{INTRODUCTION}
Recently, generation of entanglement in Cavity QED system has been
intensely paid attention to because of the motivations in the
potential applications in quantum information \cite{1,2,3} and
computation processing \cite{4,5}. The realization of atomic
entangled states in Cavity QED turns out to be feasible and
fascinating since the experimental realization of single cold
atoms in real-time Cavity QED \cite{6}. A number of schemes of
generating entanglement between cold atoms have been put forward
to realize quantum teleportation \cite{7,8,9} and swapping
\cite{10,11}. In current investigated cold-atom schemes, atoms are
trapped in optical cavities or magneto-optical traps (MOT) so that
they can be connected through exchanging single photons with
cavity (in the large interatomic separation regime)
\cite{12,13,14} or through resonant dipole-dipole interaction
(RDDI) (in the small interatomic separation regime)\cite{15,16},
and then be strongly entangled at a special interacting time.
Furthermore, in case of weak atom-field coupling and large
detuning between atomic transition and cavity frequency, even in
the large interatomic separation regime, the Rabi oscillation of
atom-field can be effectively treated as atomic RDDI which can
certainly induce diatom entanglement \cite{17}. So, in an assumed
optical cavity, the atomic entangled states are generated from the
collective contribution of Rabi coupling (RC) and RDDI. The
competition between them leads to our optimal choice of the
effective model. It has been demonstrated that, in case of diatom
scheme, the asymmetric atom-field position-dependent RC
($g_{1}\neq g_{2}$ with $g_{1(2)}$ the Rabi frequencies) depresses
the fidelity of the maximal entanglement (ME)\cite{12}. In fact,
if maximally asymmetric atom-field RC (MARC) emerges, as
$\frac{g_{1}g_{2}}{g_{1}^2+g_{2}^2}=0$, Rabi frequency can never
induce any atomic entanglement. While, under this circumstance, if
we involve the RDDI between atoms, the entanglement situation can
be different even when the RDDI is weak enough. In this paper, we
investigate such an assumed experimental situation. We will show
how to generate MARC in a Cavity QED system and what is the diatom
entanglement under various competition between RC and RDDI in it.
\section{Model Description}
Here we consider a system constituted by two-level cold atoms 1
and 2 that couple to a single mode electromagnetic cavity field
which is assumed to be a Gaussian mode profile, as is shown in
Fig. 1. The whole experimental apparatus is described as follow:
Two super-polished spherical mirrors of radius of curvature $1cm$
constructed a cavity of length $1um$ with cavity waist $w_{0}\sim
4\mu m$. In this cavity, the maximum atom-field coupling
coefficient that only occurs at an antinode of the cavity field
mode is $g_{0}\simeq 400MHz$. We choose two atomic levels are
$6S_{1/2}$ and $6P_{3/2}$ of Cesium atoms with corresponding
transition wavelength $\lambda \doteq 850nm$. We denote the
interatomic separation as $R$ and the wave vector of the atomic
emitted photon as $k_{0}$.
\begin{figure}
\epsfig{file=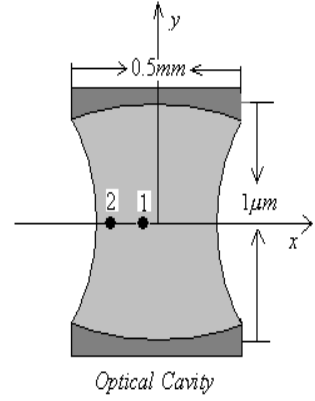, width=5cm, height=4cm, bbllx=1, bblly=0,
bburx=227 ,bbury=140}\caption{{\protect\footnotesize {Schematic
diagram for proposed model. One atom is trapped in an optical
cavity, the other is located not far away beside the former.}}}
\end{figure}

In the rotating-wave approximation (RWA), and in the interaction
picture in case of resonant coupling, the Hamiltonian governing
this system reads
\begin{eqnarray}
H=\sum\limits_{i}g_{i}(a\sigma _{i}^{+}+h.c.)+\Gamma (\sigma
_{1}^{+}\sigma _{2}^{-}+h.c)
\end{eqnarray}
where $\sigma _{1}^{z}$, $\sigma _{2}^{z}$, $\sigma _{1}^{\pm }$
and $\sigma _{2}^{\pm }$ are spin operators and raising (lowering)
operators of atom $1$ and $2$ respectively, $a^{+}$ ($a$) is the
creation (annihilation) operator of field, $g_{1}$ ($g_{2}$) is
the coupling strength of atom 1 (2) to field respectively, while
$\Gamma $ is the coefficient of atom-atom RDDI which is included
when the interatomic separation $R$ is not much two large
\cite{15,16}. It has been noted that both $g_{i}$ and $\Gamma$ are
position-dependent variables for Cavity QED theoretically and
experimentally. So that $g(r)=g_{0}cos(2\pi
x/\lambda)exp[-(y^2+z^2)/w_{0}^2]$ \cite{18}, $w_{0}$ is the
length of the cavity waist, and $\Gamma$ has a complicated form
that depends on $\frac{1}{R}$, $\frac{1}{R^2}$, and
$\frac{1}{R^3}$. The completely solving of such a system is
possible but complicated. Here we consider a physically possible
and feasible circumstance, where two atoms enter the cavity in
turn, the separation between them is in the order of cavity waist.
We also assume the atomic velocity is much small so that the
typical atomic transit time scale is much larger than the
interacting time scale. So the Hamiltonian in Equ. (1) is
time-independent. The solving of this circumstance is much more
simply and neatly. We assume atom 2 is located in a position about
$x_{2}\simeq -5w_{0}$, atom 1 is trapped in the region not far
from the antinode of the field, so that $-2w_{0}\leq x_{1}\leq
2w_{0}$. As a result, $g_{2}\sim 10^{-10}g_{1}$. This
approximation leads to a MARC Hamiltonian
\begin{eqnarray}
H=g_{1}(a\sigma _{1}^{+}+h.c.)+\Gamma (\sigma _{1}^{+}\sigma
_{2}^{-}+h.c)
\end{eqnarray}

In the invariant sub-space of the global system, for the system of
only a single photon shared by atom pairs and cavity field, the
Hamiltonian can be written on a set of complete basis $\left|
g,g,1\right\rangle $, $\left| e,g,0\right\rangle $, $\left|
g,e,0\right\rangle $ as
\begin{eqnarray}
H=\left(\begin{array}{cccc}
0 & g_{1} & 0 \\
g_{1} & 0 & \Gamma \\
0 & \Gamma & 0 \\
\end{array}\right)
\end{eqnarray}

The eigenvalues of this Hamiltonian can be obtained as $E_{1}=0$,
$E_{2,3}=\pm g_{1}\sqrt{1+\gamma^2}$, where
$\gamma=\frac{g_{1}}{\Gamma}$. The corresponding eigenvectors are
obtained as
\begin{eqnarray}
\left| \phi _{1}\right\rangle =\xi _{1}[\gamma\left|
g,g,1\right\rangle +\left|
g,e,0\right\rangle ] \nonumber\\
\left| \phi _{2,3}\right\rangle =\xi _{2,3}[\frac{1}{\gamma}\left|
g,g,1\right\rangle+\left| g,e,0\right\rangle\nonumber \\ \pm
\sqrt{1+\frac{1}{\gamma^{2}}}\left|e,g,0\right\rangle ]
\end{eqnarray}
with $\xi _{i}$ the normalized factor. The first eigenstate
$\left| \phi _{1}\right\rangle$ with corresponding zero eigenvalue
, describing atom one fixed on its ground state, can be seen as a
"dark state".

For a given initial state $\left| \psi (0)\right\rangle$ of
system, we can obtain the evolved dressed state $\left| \psi
(t)\right\rangle$ of system which can be expanded as a
superposition of eigenstates $\left| \phi _{i}\right\rangle$
\begin{eqnarray}
 \left| \psi (t)\right\rangle=\sum\limits_{i}C_{i}(t)\left| \phi _{i}\right\rangle
\end{eqnarray}
The coefficients $C_{i}(t)$ are obtained by solving
Schr\"{o}dinger Equation, so that
$C_{i}(t)=C_{i}(0)e^{-iE_{i}t/\hbar }$ with $C_{i}(0)$ determined
by initial conditions. After a sub-Doppler and evaporating cooling
suffering, atom 2 are reasonably on their ground state, i.e.
$6S_{1/2}$. While, atom 1 might in a superposition of its ground
state and excited state due to the strong correlation between it
and cavity field. The resulting initial system state is $\left|
\psi (0)\right\rangle =\left| g\right\rangle _{2}\otimes (\alpha
\left| g\right\rangle _{1}\left| 1\right\rangle +\beta \left|
e\right\rangle _{1}\left| 0\right\rangle )$. The system state is
then given by
\begin{eqnarray}
\left| \psi (t)\right\rangle =(-\mu \alpha -\nu \beta )\left| \phi
_{1}\right\rangle +[(\mu \alpha +\nu \beta )/2\nonumber \\-
\frac{\nu}{\mu} \gamma\beta]e^{-iE_{2}t}\left| \phi
_{2}\right\rangle +(\mu \alpha +\nu \beta )/2e^{-iE_{3}t}\left|
\phi _{3}\right\rangle
\end{eqnarray}
where $\mu =\frac{\gamma}{(1+\gamma) ^{2}},\nu
=\frac{\gamma}{(1+\gamma) ^{\frac{3}{2}}}$.

The reduced density matrix of diatom is obtained by tracing over
the field variables of system density matrix $\rho (t)=\left| \psi
(t)\right\rangle \left\langle \psi (t)\right| $, that is $\rho
_{atom}(t)=Tr_{f}\rho (t)=\sum\limits_{n}\left\langle n\right|
\rho (t)\left| n\right\rangle $. From Equ. (6), the atomic density
matrix is
\begin{eqnarray}
\rho _{atom}(t)=\left(\begin{array}{cccc}
D & 0 & 0 & 0 \\
0 & B & E & 0 \\
0 & E & C & 0 \\
0 & 0 & 0 & 0 \\
\end{array}\right)
\end{eqnarray}
where $E^2=B*C$. The off-diagonal element $E$ describing the
correlation between atomic transition $\left| e\right\rangle
_{1(2)}\rightarrow \left| g\right\rangle _{1(2)}$ and $\left|
g\right\rangle _{2(1)}\rightarrow \left| e\right\rangle _{2(1)}$
represents the strength of the entangled state for two atoms. In
the next section, we will discuss the entanglement induced by the
RDDI between atoms.
\section{Diatom entanglement nature under cavity field}
Wootters Concurrence, which has been proved to be effective in
measuring the entanglement of two qubits, is defined as\cite{19}
\begin{eqnarray}
C(\rho )=\max \{0,\lambda _{1}-\lambda _{2}-\lambda _{3}-\lambda
_{4}\},
\end{eqnarray}
where $\lambda _{i}$ are four non-negative square roots of the
eigenvalues of the non-Hermitian matrix $\rho (\sigma _{y}\otimes
\sigma _{y})\rho ^{\ast }(\sigma _{y}\otimes \sigma _{y})$ in
decreasing order. In dealing with this model, the Concurrence is
simply determined by several density matrix elements since most of
the off-diagonal elements are eliminated due to the adiabatic
evolution. In our case, the Concurrence can be proved to be
exactly twice the absolute none-zero off-diagonal element as
$C(\rho(t) )=2|E(t)|$.

For large interatomic separation $R$, which corresponding the
far-zone case with $Rk_{0}\sim 10^2\gg 1$, the dipole-dipole
interaction strength can be estimated as $\Gamma\sim \omega _{a}
a_{0}^3k_{0}^2/R$\cite{15,20}, where $c$ is light velocity,
$a_{0}$ is the atomic excited state radius. This emerges the limit
of the RDDI strength as $\Gamma\sim 10^5 Hz\leq 10^{-2}g_{1}$. Our
further assumption is the atom that earlier enters the cavity is
now also in its ground state, i.e. $\beta =0$ in Equ. 6.
(Certainly, this assumption leads to the half possibility of the
successively entangling two atoms.)
\begin{figure}
\epsfig{file=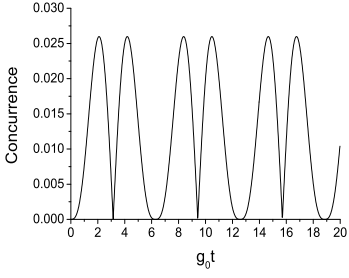, width=5cm, height=4cm,bbllx=8, bblly=9,
bburx=320 ,bbury=243} \caption{{\protect\footnotesize {The
evolution of diatom entanglement versus time, where the location
of atom 1 is $x_{1}=-2w_{0}$}}}
\end{figure}

The diatom entanglement shows two-peak split structure. Each peak
corresponds to a PE which occurs at $t=\frac{(3m\pm 1)\pi}{3}
\frac{1}{\sqrt{g^2+\Gamma^2}}$ with $m=1,3,5\cdots$ (see Fig. 2).
The increase of time leads to an entanglement period that depends
on the Rabi frequency of atoms and RDDI strength. Both the amount
and period of PE are position-dependent. When the initial location
of atom 1 departs from atom 2 gradually, they decay exponentially
and fall to their minimum at the center of the cavity, see Fig. 3
and Fig. 4. These can be understood since Rabi frequency increases
exponentially in this process and reaches its maximum at the
center of the cavity, while RDDI strength decreases polynomially.
Similarly, as the initial location of atom 1 increases along $x$
axis gradually from the center of cavity, the amount and period of
PE can be recovered since Rabi frequency decreases now.
\begin{figure}
\epsfig{file=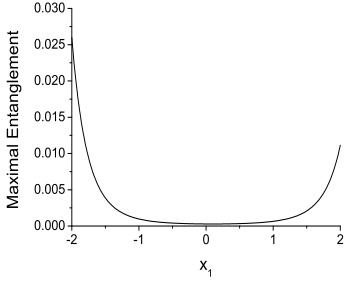, width=5cm, height=4cm,bbllx=7, bblly=10,
bburx=320 ,bbury=237} \caption{{\protect\footnotesize {The diatom
PE occurs at $t=\frac{2\pi}{3} \frac{1}{\sqrt{g^2+\Gamma^2}}$
versus the location of atom 1 $x_{1}$ (in unit of cavity waist
$w_{0}$).}}}
\end{figure}

In fact, from Equ. (6) we can see that, the amplitude of the
entanglement is only determined by the ratio of RDDI and Rabi
frequency of atom 1. In the region of $w_{0}\leq
\left|x_{1}\right|\leq 2w_{0}$, the ratio (namely $\gamma$) is in
the order of $10^{-3}$. In this region, the effect of RDDI can not
be neglected since the PE is distinct, even the interatomic
separation can approach $7w_{0}$.

While, in the region of $\left|x_{1}\right|\leq w_{0}$ (which
corresponds to a fast-oscillating regime), the coupling between
atom 1 and cavity field is so much strong that the ratio $\gamma
\approx0$. The Hamiltonian in Equ. (2) can be adiabatically
written as
\begin{eqnarray}
H_{eff}=g_{1}(a\sigma _{1}^{+}+h.c.)+\frac{2\sqrt{2}\Gamma
^{2}}{g_{1}}(\sigma _{1}^{z}\sigma _{2}^{+}\sigma _{2}^{-}-\sigma
_{2}^{z}\sigma _{1}^{+}\sigma _{1}^{-})
\end{eqnarray}
The system described by this Hamiltonian can never generate
entanglement since the diatom coupling part in Equ. (9) is
diagonalized.
\begin{figure}
\epsfig{file=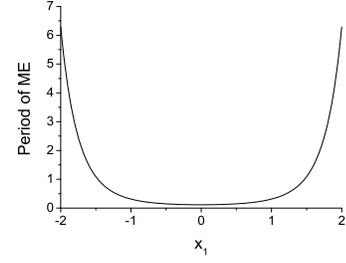, width=5cm, height=4cm}
\caption{{\protect\footnotesize {The period of PE (in unit of
inverse of $g_{0}$) versus the location of atom 1 (in unit of
cavity waist $w_{0}$).}}}
\end{figure}

In order to give an overall description of the diatom
entanglement, we mesh the concurrence under the position of atom 1
and time in Fig. 5. From this figure, we can clearly get the hint
of two-peak split structure and the period of PE.

All the results reveal that, in the weak Rabi coupling region, the
amount and the period of PE is larger than that in other region.
Since the participation of atom-field interaction results in
diatom mixed state which can not be an exactly Bell-state, the
quality and quantity of PE is strongly depressed.

\begin{figure}
\epsfig{file=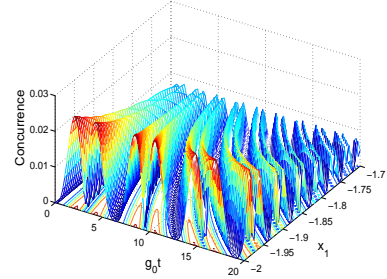, width=5cm, height=4cm}
\caption{{\protect\footnotesize {The evolution of diatom
entanglement versus time and the location of atom 1 (in unit of
cavity waist $w_{0}$).}}}
\end{figure}

\section{Conclusion}
We investigate a diatom entanglement case for maximally asymmetric
Rabi coupling in Cavity QED. The characterized apparatus
parameters are constituted for experimental Optical Cavity. The
entanglement situation can be analyzed bases on the ratio of
position-dependent RDDI strength and Rabi frequency of atom 1. We
concretely point out the amplitude (which is a two-peak split
structure) and the period of the entanglement in two regions. In
the region of relative large ratio, where the RDDI strength can be
in the order much larger than $10^{-3}g$, diatom entanglement can
be distinctly generated. We do not involve the motion of
center-of-mass of two atoms. While, if the atomic velocity is
large enough and the typical transit time scale can compare with
the interacting time scale of the system, the effect of the motion
on the entanglement must be included. And the Hamiltonian will be
time-dependent one. Additionally, our MARC approximation should be
still effective for wider initial location of atom 1. Even if atom
1 is initially located at $x_{1}(0)=\pm 3w_{0}$, where the RDDI
strength would be in the same order of Rabi frequency and the
ratio $\gamma$ can approach $1.2$, the validity of this
approximation keeps well since $g_{1}\sim \Gamma \sim 10^7g_{2}$.
It should be pointed out that, actually, an underlying trouble for
the fidelity of the results here may be the cavity dissipation and
the atomic spontaneous emission that can lead to an exponential
decay to the excited state population. But, in consideration of
the high-Q regime of the cavity and the fact of cavity compressing
atomic spontaneous emission for not two large interatomic
separation \cite{21}, in a short-time range of the evolution, the
trouble may be faint and nonsignificant.

\section{Acknowledgments}
This work was supported by NSF of China, under grant NO. 10347103,
10305002 and 60472017.

\end{document}